\documentclass{svmult}
\usepackage{url}
\usepackage{epsfig}
\usepackage{algorithm}
\usepackage[noend]{algorithmic}

\usepackage{algorithm}
\usepackage[noend]{algorithmic}
\usepackage{graphicx}
\usepackage{subfigure}
\usepackage{amssymb}
\usepackage{amsmath}
\usepackage{stmaryrd}

\begin{document}

\author{Walter Quattrociocchi, Rosaria Conte, Elena Lodi}

\title{Opinions within Media, Power and Gossip}

\maketitle

\begin{abstract}
Despite the increasing diffusion of the Internet technology, TV remains the principal medium of communication. People's perceptions, knowledge, beliefs and opinions about matter of facts  get (in)formed through the information reported on by the mass-media. 

However, a single source of information (and consensus) could be a potential cause of anomalies in the structure and evolution of a society.

Hence, as the information available (and the way it is reported) is fundamental for our perceptions and opinions, the definition of conditions allowing for a good information to be disseminated is a pressing challenge. 
In this paper starting from a report on the last Italian political campaign in 2008, we derive a socio-cognitive computational model of opinion dynamics where agents get informed by different sources of information. Then, a what-if analysis, performed trough simulations on the model's parameters space, is shown. 
In particular, the scenario implemented includes three main streams of information acquisition, differing in both the contents and the perceived reliability of the messages spread. Agents' internal opinion is updated either by accessing one of the information sources, namely media and experts, or by exchanging information with one another. They are also endowed with cognitive mechanisms to accept, reject or partially consider the acquired information.
\end{abstract}


\section{Introduction}
Despite the increasing diffusion of the Internet technology, traditional media -- e.g., news papers and TV - remain the principal instruments for the information diffusion.  People's perceptions, knowledge, beliefs and opinions about the world and its evolution, get (in)formed and modulated through the information reported on by the mass-media.  An unbalanced  distribution of the power on this basin of information (and consensus) could lead to anomalies in the structure and the evolution of society.
In this paper we focus upon a) the possible effects that an anomalous distribution (and use) of the mediatic power could {\bf potentially} bear on the public opinion b) the conditions allowing for a good dissemination of the information.

Since the early thirties (\cite{Sherif88,Asch55}) the informal study of social influence has produced abundant evidence of structural factors affecting people's beliefs. In social life, agents are exposed to different communication systems interacting with one another, ranging from one-to-many information transmission typical of traditional broadcasting media, to one-to-one systems characterizing the new media, but including several intermediate modalities. How do they interact? Which one is most likely to exercise the strongest influence on agents' opinions? Despite the importance social scientists attribute to the role of persuasive communication (think of the Hovland school of persuasion), different communication systems have rarely been compared under natural conditions, and even less in artificial experiments. Based on social impact theory (\cite{nowak90}) recent simulation-based studies of opinion dynamics \cite{Staufer2007,Lorenz07,niguez09opinion,Castellano2007,Hu09,brunetti2010,AQ2010a,QuattrociocchiPC09} observe how numerically defined opinions spread and aggregate over a given population as a function of the distance among the values agents assign to them. Within these studies, however, the process of communication among the agents is not explicitly addressed. Plunged into the same network, agents are assumed to exchange opinions as a function of the distance between them: the lower this is, the more the agents are inclined to converge. In this paper, the role of different forms of communication in opinion dynamics is addressed with the help of agent based simulation. 

The design of the computational model has been derived by a survey reporting on the relationship between information delivered by the media and the social perceptions' dynamics during the Italian political campaign in 2008. 
In our model, opinions are numerically defined on one parameter that stands for the certainty with which agents hold them; as in the bounded confidence model (\cite{amblard01}), agents are assumed to exchange their opinions based on the distance between them.  However, we introduced three modifications over the preceding works on opinion dynamics: agents (a) share two relatively independent opinions, (b) are exposed to different forms of communication, namely one-to-many and one-to-one, which will be characterized later on in the paper on a number of dimensions,  and (c) receive inputs from two distinct sources of information, expert and non-expert. 

The former modification is suggested by the necessity to make the scenario more realistic. 
The latter modifications instead are needed to address a topical question. It is a common opinion nowadays that the new media play a positive role in the control and improvement of the quality of information circulating in a system. To what extent is this opinion backed up by existing evidence? Comparing the effects of peer-to-peer communication Vs traditional media on the diffusion of opinions and observing their interaction, we aim to investigate whether the former system is bound to amplify the effects of the latter, or can exercise a relatively independent influence on information quality. To clearly distinguish the effects of the two systems, we have introduced the distinction between the expert and the non-expert source. To what extent will the experts affect existing opinions? How much expert-driven information must be accessed through the new media for the average quality of information circulating in the population to increase?

\section{Grounding the model}

\subsection{Media in the Italian Case}
Due the abstract and complex nature of the phenomenon, capturing the interplay between information delivered by the media and their effects on social perceptions in a consistent socio-cognitive computational model is absolutely not trivial.
The theory of “agenda-setting” presented in \cite{Maxwell} shows that the frequency of information delivered by the media correlated positively with its consequent perceived importance. The issues reported on by central media are the most debated and are perceived as priorities to be solved by policy-makers. Hence, by ``bombing'' the audience with the same news the media could {\bf potentially} impact and modify the public opinion. (Un)Fortunately the actual Italian scenario provides a very ``luxurious'' real context to analyze such a sophisticated matter in depth. In the following we show charts from a survey named "Security in Italy: Meanings, Image and Reality''  (\cite{IndagineDemosUnipolis})  performed during the last Italian political campaign (2008). Such a survey shows that the quality of the reported information is often lower than desirable. This should not come as a surprise, if we compare the communication from central media to market asymmetry \cite{akelrof70}. Under asymmetry, information, as any other product, is bound to become a “lemon”, i.e. a corrupted and useless good.   

\begin{figure}[h!]
 \centering
 \includegraphics[width=100mm]{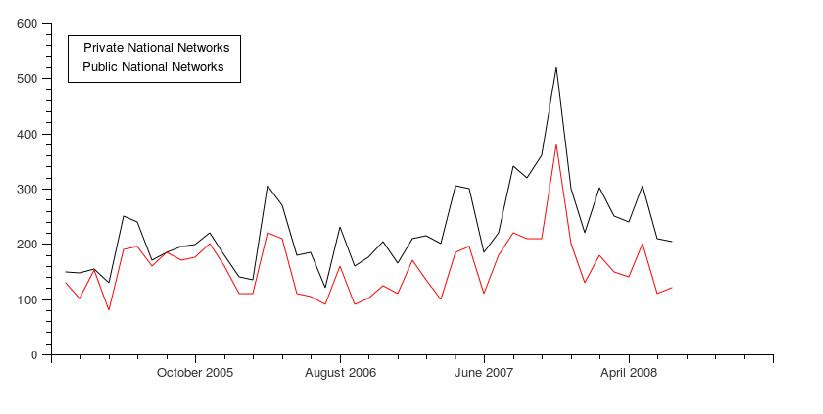}
 \caption{The trend of news related to common crimes delivered by the private (black) and public (red) networks }
 \label{fig:news}
\end{figure}

\begin{figure}[h!]
 \centering
 \includegraphics[width=100mm]{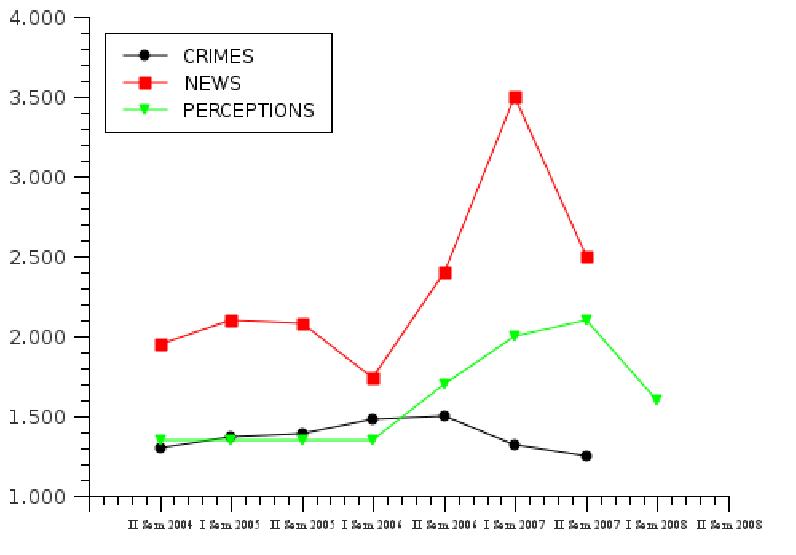}
 \caption{The trend of the news reported on by media about common crimes in red, the trend of social perception about crimes in green, and the real trend of common crimes in black}
 \label{fig:perception}
\end{figure}

Figure \ref{fig:news} shows the number of news regarding common crimes during four years comparing the news reported on respectively by private (in black) and by public national networks (in red). Notice that the highest peaks of both curves coincide with the political campaign period. 
Figure \ref{fig:perception} shows the real number of crimes (in black), the media reports on security matters (in red), and the shared social perceptions of the sample investigated within the survey (in green). The media reports about security are shown to be almost perfectly complementary to the real trend of crimes, as provided by the Ministry of Domestic Affairs.
As figures show, both networks - the private more than the public - injected informational lemons into society during the last Italian campaign.  
The person holding the private network (and indirectly controlling a part of the public network service) and the candidate proposing security as the principal Italian problem, coincide.
He won the election.
Which kind of role has been played by such a strategy of communication, supported by the media   reports, on the public agenda and on the elections' results?

\subsection{Within Different Communication Paradigms}
Information is acquired both through communication among agents and from the centralized media. One-to-one and one-to-many communication have been compared by marketing scientists (\cite{Deighton95}) as well as cultural evolutionary scientists (\cite{cavalli-sforza86}), and shown to have different although balanced effects.
As reported in \cite{eRep}, according to classic communication models (see \cite{Hoadley99}), peer to peer (P2P) communication is a step-wise, asynchronous process that requires a more or less lengthy temporal extension: it starts at one point in time and takes effect after a certain amount of steps, and in each of them it is iterated. On the contrary, centralized or broadcast communication takes place at once. Whereas broadcast communication is a one-to-many process (\cite{Harris04}), P2P communication reaches a smaller audience.
As summarized in Table \ref{tab:pvsbc}, two complementary patterns of properties emerge. Each pattern allows different expected performances. P2P systems are less efficient and more liable to corruption, although more interactive and controllable. P2P can either be proactive or reactive, whilst broadcast (BC) communication generally is only proactive. The former is based upon and aimed at reciprocating information, whereas broadcast communication can hardly be expected to be reciprocated (it is often institutional). P2P is spontaneous and based upon acquaintanceship or familiarity networks, while BC communication, which is based upon other types of networks, is often facilitated or allowed by the sharing of new technologies. 

Which further consequences can we expect from either system? How deep is their respective influence on the population? How do they interact, when insisting on the same population? Are they interdependent, or is there a dominance of one system on the other, and if so, which one is more influential? Finally, what is the effect of their interaction on information quality?

\begin{center}
\begin{table}
\centering
\label{tab:pvsbc}
 \begin{tabular}{|c|c|}
\hline 
Peer-to-Peer Communication (P2P) & Broadcast Communication (BC)\tabularnewline
\hline
\hline 
one to one/few & one to many\tabularnewline
\hline 
asynchronous & synchronous\tabularnewline
\hline 
step wise & at once\tabularnewline
\hline 
sensitive to temporal extention  & not affected by temporal extention\tabularnewline
\hline 
interactive & proactive\tabularnewline
\hline 
based upon reciprocation & no reciprocation\tabularnewline
\hline 
based upon familiarity networks & not affected by the network topology\tabularnewline
\hline 
\end{tabular}
\end{table}
\end{center}

\section{Context}

\subsection{Related Works}
The effect of communication on opinion formation has been addressed by different disciplines from within the social and the computational sciences, as well as the complexity science. 
Social scientists focus on polarization, i.e. the concentration of opinions by means of interaction, as one main effect of the "social influence'' \cite{festinger50}. Social psychology offers an extensive literature on attitude change models, as reviewed by \cite{Mason07}. 
Most influential in social psychology is the ``The Social Impact Theory'' \cite{Latane90}, according to which the amount of influence depends on the distance, number, and strength (i.e., persuasiveness) of influence sources. As stated in (\cite{Castellano2007}), an important variable, poorly controlled in current studies, is structure topology. Interactions are invariably assumed as either all-to-all or based on a spatial regular location (lattice), while more realistic scenarios are ignored. The most popular model applied to the aggregation of opinions is the bounded confidence model, presented in \cite{amblard01}.
Much like previous studies (\cite{Lorenz2008,Galam08}), in this paper agents exchanging information are modeled as likely to adjust their opinions only if the preceding and the received information are close enough to each other. Such an aspect is modeled by introducing a real number $\epsilon$, which stands for tolerance or uncertainty (\cite{Castellano2007}) such that an agent with an opinion $x$ interacts only with agents whose opinions is in the interval $] x - \epsilon ,  x + \epsilon[$.
In our previous works \cite{quattrociocchi2008}, \cite{quattrociocchi2008b}, \cite{quattrociocchi2009b}, \cite{Balke08} and \cite{eRep} we investigated the role of communication systems on agents' perceptions, by means of multi-agent based simulations, when informational cheating occurs.

The model we present in this paper, which preliminary results have been introduced in \cite{quattrociocchi2010d,quattrociocchi2010e}, extends the bounded confidence model by providing agents with two, instead of one, conflicting and independent values representing their opinions about, say, welfare and security. 
Furthermore, in our model agents resort to two additional sources of information, external to the social network, aimed at representing experts and media.

\subsection{Research Questions}

In our previous work \cite{quattrociocchi2010d} the focus was on the interplay between institutional broadcasting and P2P communication. 

The correlation within the frequency of information delivered by the media and their consequent effects on social perceptions, namely agenda-setting, has been theorized in 1972 by McCombs et al.
With the advent of the Internet new and indipendent sources of information are available to users. Furthermore, an individual on the Internet can select the sources which is closer to his/her vision about matter of facts or having direct access to experts.
In this paper, we will assume that, when accessing Internet, the agents are quite confident in the truth-value of the acquired information. Of course, this assumption is somewhat arbitrary, but what matters here is the source of influence rather than the way it is found out.

In our model, agents are exposed to a) the conventional media, repeating the same message at each time step,  b) the new media, and c) to the information circulating within the neighborhood. 

Hence, the follow-up research questions addressed within the present work: (a) what would happen to agents' opinions if both conventional and new media were confronted with an additional P2P-based source of information highly trusted by the agents? (b) Which is the role of the white-zone, namely the percentage of agents that are reached neither by media nor on the Internet? Furthermore, how many experts are needed in a network to reduce the information asymmetry between agents and conventional media, or, to put it more explicitly,  how many experts are needed to contrast possible informational cheating spread within the system? (d) How does the interaction topology, i.e. the network structure, affect the information diffusion and the dynamics of opinions? In this paper we address the former three sets of questions, leaving the last one for future works.

\section{Preliminaries}

In the simulated system, traditional media send out messages in broadcasting to a variable percentage of the population, while members communicate with neighbors and are thus exposed to an additional source of information which they consider highly reliable, the experts.

In this section the basic definition related with our model and the simulation results are provided.

\subsection{The Interaction Topology}

In order to design a realistic model our agents are plunged in a network incorporating real world networks' properties. 
Real world networks are “scale-free”, in the sense that their node degree distributions follow a power-law that is not affected by the size of the network.
Let $N$ be a connected graph in which each node $v$ has a number of $k$ originating links following a power law distribution $P(k)\sim k^{-y}$.
We generate a scale free network by progressively adding nodes on a previously existing network and then introducing links to the existing nodes following the so called ``preferential attachment'' mechanism \cite{Barabasi99}. The construction strategy of the algorithm aims at maintaining the link probability between any couple of nodes proportional to the number of existing links $k_{i}$ already connected to the selected node.
\subsection{The General Bounded Confidence}

As mentioned above, the most famous model of opinion dynamics based upon bounded confidence is the one developed by Deffuant et al. in \cite{amblard01}. The model can be explained as an asynchronous game in a distributed environment where the nodes $v$ of the network $N$, i.e., the agents, interact by exchanging their opinions.
For instance, consider the iterative process over $N$, where each element $v \in N$ updates its internal state at each time step by comparing its opinion with the information circulating in the neighborhood. From a theoretical point of view, this model presents a number of oversimplifications. In particular, it assumes that low distance is the only determinant of opinion update. Other aspects, for example, (a) the degree of certainty on one's opinion and (b) the extent to which it is shared by others are ignored. The second aspect will be addressed in future works. For the time being, we limit ourselves to extend the original model to a slightly more complex situation in which agents have a generic subjective disposition to accept others' beliefs, and hold two independent opinions. Hence, given two agents $x$ and $y$ exchanging their opinions $v(x)$ and $v(y)$ the entities' internal states are updated by applying the following algorithm:

\begin{center}
\begin{algorithm}[h!]
  \caption{(BCM) Bounded confidence model}
\begin{algorithmic}[h!]
\IF {$\mid v(x)-v(y)| \leq t$}
        \STATE  $v(x) \gets v(x) + m (v(y)-v(x))$
\ENDIF
\end{algorithmic}
\end{algorithm}
\end{center}

Where $m$ is a constant that can be fixed by the user within the interval $(0..0.5]$. It represents the convergence parameter - e.g., a way to increase the convergence by dividing the distance $d = v(x)-v(y)$ between two agents' opinions in $\frac{1}{m}$ steps. 
The variable $t$ represents agents' tolerance, a threshold defining the limit under which an opinion can be accepted by an agent.

\section{The Model}
\begin{figure}[h!]
 \centering
 \includegraphics[width=210mm,bb=0 0 1244 392]{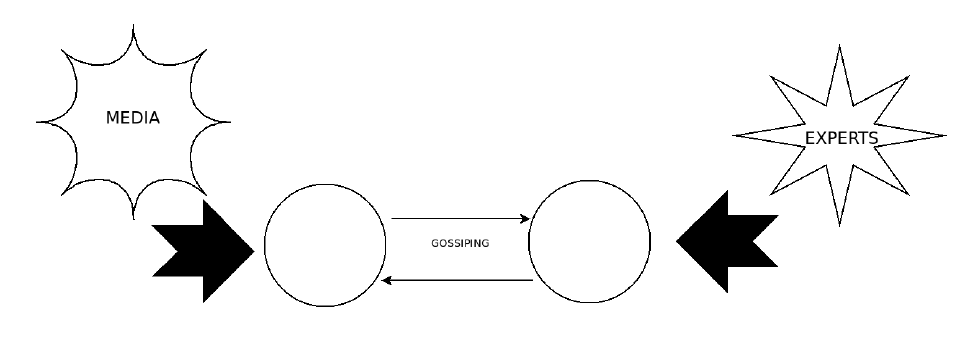}
 \caption{The Communication Model}
 \label{fig:interaction}
\end{figure}

Our model extends the model of Deffuant et al. by defining agents beliefs on two conflicting values measuring respectively the welfare and security desire. In addition, as shown in \textbf{Figure \ref{fig:interaction}} in our model agents get informed by accessing two different additional sources of information, namely {\em experts} and {\em media}.

We refer to agents accessing the former source as wise agents, while agents accessing the latter are called televiewers.  In general, to process information, agents apply the bounded confidence; on the contrary, a wise agent is more inclined to accept the information acquired by the experts.

\subsection{Entities of the Model}
In this section the agents protocol and their interactions will be introduced.

\subsubsection{Media.}
Conventional media is simulated as a special agency, reporting the same message at each simulation turn to a subset of the agents' set $V$  .  

The media agent $m\in M$ is not linked to the network and has the goal to persuade the audience that security is a matter more important than welfare. The media's reported message is denoted by the following set:

\begin{equation}
\begin{split}\left\lbrace m_{l},m_{r},V_{1}\mid0\leq m_{l}\leq1\wedge0\leq m_{r}\leq1,V_{1}\subset V\right\rbrace \end{split}
\end{equation}

where $m_{l},m_{r}$ represent the media reported values of events respectively related to welfare and security issues.

\subsubsection{Interacting Peers.}
The audience is composed by agents (i.e. the network nodes), whose shared goal is to exchange information with other agents in one's neighborhood. They also receive messages from the media and from the wise agents.

When interacting with one another, agents  $v\in V$ are provided with an internal state defined as follows:

\begin{equation}
\left\lbrace v_{l},v_{r}\mid0\leq v_{l}\leq1\wedge0\leq v_{r}\leq1\right\rbrace 
\end{equation}

respectively representing agents' beliefs about welfare and security issues. 
The closer $v_{l}$ and $v_{r}$ values are to 1, the more each debated issue (i.e. welfare and security) is considered to be important by the agent. The agents' internal state, in the protocol, corresponds to the message sent as an answer to each external request of information coming from its neighbours. 
The initial configuration of agents' opinions is set up according to a uniform random distribution of both values (i.e. security and welfare desire).

\subsubsection{Wiseagents and Televiewers.}

The set of agents $V$ is composed by two kinds of agents, each subpopulation is denoted by the source of information accessed.
Agents accessing experts will be called WiseAgents (WAs), those exposed to media will be called TeleViewers (TVs). 

WiseAgents and TeleViewers differ in the way they process the information acquired. The former are highly confident in its truth value while the latter process the media information as the information they receive from peers, i.e. by applying the bounded confidence model mechanism.

\subsection{The Model Interaction}

\subsubsection{Media and Peers.}

Agents acquire information from media agencies according to a passive protocol, by acquiring the values they send and comparing them with their previous opinions. 
Information is either accepted or not, based on the bounded confidence mechanism (\cite{amblard01}). 
The agent's opinions $v_{l},v_{r}$ and the information from the media $m_{ld},m_{rd}$ are transformed in two new agent's opinions. The function generates two new values for $v_{l},v_{r}$.
The variable $t \in \mathbb{R}$ stands for peer agents' tolerance, i.e., the subjective disposition to accept others' information. The two
guard variables $g_{l}$ and $g_{r}$ are calculated by the Boolean expression returning true if the difference between
acquired and owned information is below the tolerance threshold $t$.
The guard variables $g_{l}$ and $g_{r}$ respectively control the access to the updated values of $v_{l}$ and $v_{r}$, which is implemented on two independent opinion spaces.
The values of $v_{l}$ and $v_{r}$ are updated through the following:

\begin{center}
\[
\begin{split}v_{l}=(v_{l}+(m_{ld}(v_{l}-m_{ld}))\\
v_{r}=(v_{r}+m_{rd}(v_{r}-m_{rd}))\end{split}
\]
\par\end{center}

\subsubsection{Among Interacting Peers.}

Agents exchange information by comparing their preferences. This interaction is executed after both TeleViewers and WiseAgents receive the information by their respective sources.

Each agent communicates with the set of neighbours within a distance set to 1. We will follow  \cite{amblard01}' convention, according to which, when communication occurs between any two agents, these mix their opinions when the differences is smaller than the threshold $t$.

\section{Experiments}

The experiments design has been performed with the main aim to perform a what-if analysis, based upon simulations, of the effect of information and communication on social perceptions.

We focus on the effect of two different sources of information, reporting different (and complementary) messages to the audience, on the agents' opinions trend. In particular we stress the polarization of opinions toward one of the two main debated issues, which in the model corresponds to security and welfare.
The main goal of the media is to persuade the audience that security is more important than welfare (reporting the same message respectively fixed to 8 for security and 3 for welfare). On the contrary, agents resorting to experts (giving respectively 3 for security and 8 for welfare) consider welfare more urgent than security. In addition, experts and media differ as to the way their information is trusted by the agents. On the one hand, Wise Agents accept the information provided by the experts without applying the proviso of the bounded confidence model, meaning that the information is assumed as truthful. 
On the other hand, TeleViewers, when acquiring information from the media, adjust their beliefs according to the bounded confidence model.

\subsection{Scenario 1: Gossiping Peers}
In the first set of experiments, intended to provide the scenario with the baseline settings, we explore P2P communication mechanisms according to the bounded confidence model (\cite{amblard01}), by varying the value of tolerance, i.e. the subjective disposition of agents to accept the information received, independent of the source. The baseline experimental settings implement nine scenarios, with the number of agents set to 100, without media broadcasting and for increasing levels of tolerance (from 0.1 to 0.9 at step 0.1) for 100 simulation turns. In order to cope with the computer pseudo randomisation, the simulation is performed ten times per each scenario, and the results are then averaged.

\subsubsection{Emerging Results}

\begin{figure}[h!]
 \centering
 \includegraphics[width=70mm]{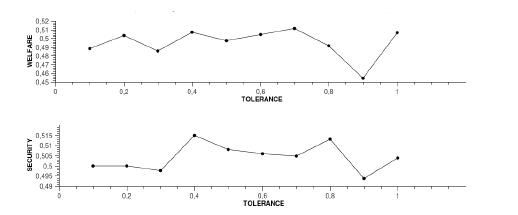}
 \caption{Interacting peers opinion dyamics for increasing level of tolerance}
 \label{fig:gossip}
\end{figure}

The effect of peer-to-peer communication on opinion dynamics is shown in Figure \ref{fig:gossip}, where both welfare and security opinions for increasing levels of tolerance
are shown. Both opinions (welfare and security) are set up randomly within the interval {\em ]0, 1[} at the beginning of each simulation. As shown in Figure \ref{fig:gossip} the trend of both opinions fluctuates around the average value of the initial distribution, meaning that, over a scale free network, the P2P communication leads to a flat distribution of both opinions about security and welfare urgency.

\subsection{Scenario 2: Adding the Media}
The second battery of experiments addresses the question what is the combined effect of the two complementary communications systems: how do they interact? In particular, does P2P communication amplify or inhibit the effect
of media? The parameters’ space has been explored for increasing numbers of
agents reached by the media broadcasting (from 0\% to 100\% with an incremental step of 10\%) and for increasing values of tolerance (from 0.1 to 0.9 with step 0.1). For each simulation ten runs were performed.

\subsubsection{Emerging Results}

\begin{figure}[h!]
 \centering
 \includegraphics[width=130mm]{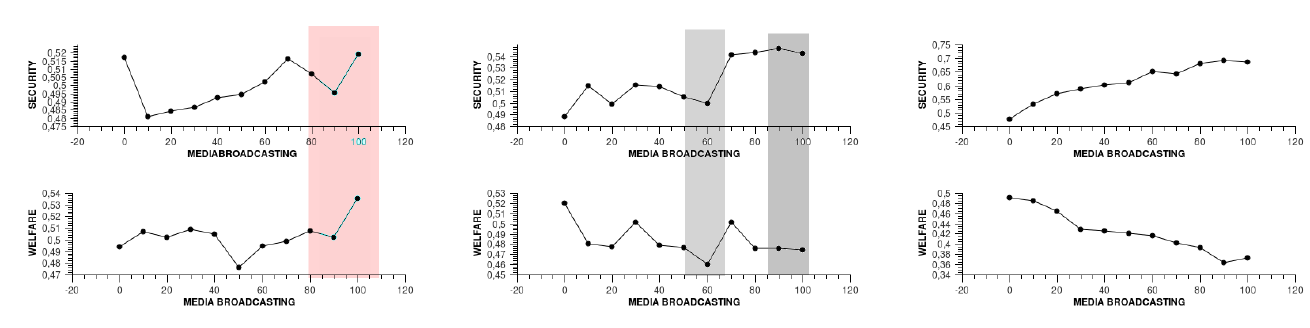}
 \caption{The effect of media message on Opinion Dyanmics for increasing levels of Tolerance (0.3,0.5,0.8)}
 \label{fig:mediaAndGossip}
\end{figure}

Opinion dynamics is based upon and mediated according to the the bounded confidence convention. 
The experimental results are presented for increasing levels of tolerance at different levels of media broadcasting. 

Figure \ref{fig:mediaAndGossip} shows the different trends of opinions under agents’ exposition to the media
broadcasting at different tolerance levels. At each turn, the central media deliver the same message, with the matters’ values respectively set to 0.8 for security and 0.3 for welfare.

As one can see in the first chart of Figure \ref{fig:mediaAndGossip}, showing the trend of agents’ opinions on security and welfare for increasing exposition to media when the tolerance is fixed to 0.1, the media affect the opinion of agents also at low levels of tolerance but with a lower impact. Notice that when the media broadcasting reaches all the agents (MB= 100\%), the values for security and welfare get closer to the values they have in absence of centralized media. 

When tolerance is low - meaning that agent are less inclined to accept others’ information as reliable - P2P inhibits
at least partially the effect of media. However, results show an interesting effect of non-linearity. As shown in the second chart, when tolerance is fixed to 0.5 and broadcasting is above 70\% (i.e. a strong majority of agents are reached by central media), the informational cheating is amplified by the P2P communication, clearly affecting the agents perceptions and driving their opinions toward the values reported on by the media. 

However, when boradcasting is below 70\% at the same level of tolerance (0.5), Peer-to-Peer communication reduces the effect of
informational cheating, allowing for more realistic information to spread. When agents are neither totally prone to accept others’ information nor completely refractory to it, P2P communication inhibits at least partially the broadcasting informational cheating. 

Tolerance at 0.5 balances informational cheating until broadcasting reaches the 40\% of the population. Instead, when tolerance is set to 0.8 (third chart of Figure \ref{fig:mediaAndGossip}), the process is linear: the average opinions on security and
welfare increasingly approximate the values reported on by the media, depending on the values of broadcasting. The higher the tolerance, the poorer the information quality: agents find no shelter against informational cheating.

\subsection{Scenario 3: Media and Experts}

The third set of scenarios aims at exploring the consequences of different percentages of agents exposed to the main streams of information. 
Hence, each scenario is characterized by an increasing degree of exposure to conventional media and by a decreasing number of experts.

As one can see from the experimental settings listed in details in Table \ref{tab:expII}, the population is composed by 100 agents (NA) endowed with the ability to process the information through a variable level of confidence (TOL). A subset of agents (WAs) can access a different source of information with values different from the ones reported on by the media (TVs).
Each scenario has been simulated in ten runs.
In short these experiments are aimed at understanding the mutual effects of different information (with complementary values) delivered by different agencies.

\begin{table}[h!]
\centering
\caption{Experiments Settings }
\begin{tabular}{|c|c|c|l|} \hline
NA&MB&WAs&TOL\\ \hline
100 & 0& 100&0.2,0.5 and 0.8\\ \hline
100 & 10& 90&0.2,0.5 and 0.8\\ \hline
100 & 20& 80&0.2,0.5 and 0.8\\ \hline
100 & 30& 70&0.2,0.5 and 0.8\\ \hline
100 & 40& 60&0.2,0.5 and 0.8\\ \hline
100 & 50& 50&0.2,0.5 and 0.8\\ \hline
100 & 60& 40&0.2,0.5 and 0.8\\ \hline
100 & 70& 30&0.2,0.5 and 0.8\\ \hline
100 & 80& 20&0.2,0.5 and 0.8\\ \hline
100 & 90& 10&0.2,0.5 and 0.8\\ \hline
100 & 100& 0&0.2,0.5 and 0.8\\ \hline
\end{tabular}
\label{tab:expII}
\end{table}

\subsubsection{Emerging Results}

\begin{figure}[h!]
 \centering
 \subfigure[Tolerance: 0.2]
   {\includegraphics[width=50mm]{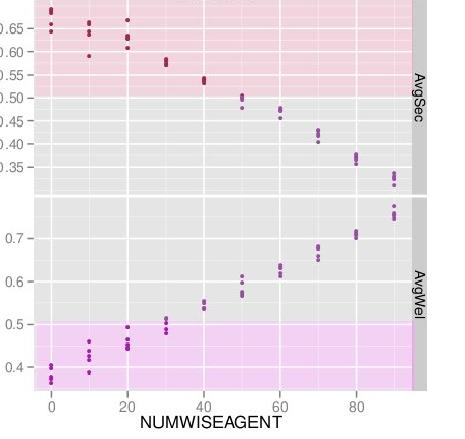} \label{fig:IIWISE02}} 
 \hspace{1mm}
 \subfigure[Tolerance: 0.5]
  {\includegraphics[width=50mm]{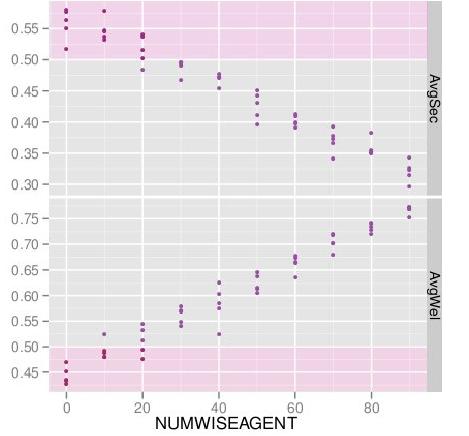} \label{fig:IIWISE05}} 
\hspace{1mm}
\subfigure[Tolerance: 0.8]
  {\includegraphics[width=60mm]{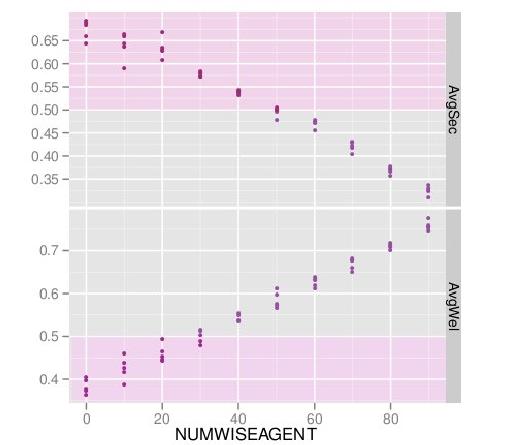} \label{fig:IIWISE08}} 
 \caption{Opinion Trend under Peer Pressure without Black Zone}
\label{fig:expIIP}
\end{figure}

The results emerging from the first battery of experiments are shown in \textbf{Figure \ref{fig:expIIP}} where the aggregate values of both agents' opinions on security and welfare matters, for increasing presence of WAs, are reported.
The figures report the opinion trends for different levels of tolerance (0.2, 0.5, 0.8). 

All figures show the same effect: the higher the presence of WAs the more the agents' opinions converge toward the values reported on by the experts (3 for security and 8 for welfare). On the contrary, namely when the media reach the largest amount of the total population (and WAs are on the lowest level), no convergence can be observed but only the reduction of the distances among values. 
In our view this phenomenon indicates that, consistent with our previous results, the efficacy of traditional media are sensitive to the effect of peer-to-peer communication. By definition WAs do not process the information with the bounded confidence mechanism, thus the tolerance does not affect opinion adjustment.

Notice the extent to which the opinion trend passes the line of the average value (i.e. $0.5$). The WAs effect is relevant when the global tolerance is fixed to 2: the 30\% of WAs is sufficient to invert the effect on mass opinions of the 70\% of TWs.
Furthermore when tolerance is higher (5), the number of WAs needed to inhibit media is smaller (20\%). 
One may expect that tolerance plays a fundamental role in opinions' convergence toward the media or the WAs values.
Results indicate that the high level of confidence in buying others' information has a side effect: WAs are less efficient when tolerance is high, meaning that if uncertainty is strong agents are inclined to accept any information as truthful, even that provided by the media.

\subsection{Scenario 4: Media, Internet and Uninformed Agents}

The last scenario is similar to the one described in the previous section, except for the different distribution of WAs and TVs. 
In the preceding experiment the total amount of the population is either reached by the media, or resorts to experts. 
Peer-to-peer communication inhibits the spreading of information from the media, but what would happen if a given percentage of agents is reached neither by the media nor by the experts? In other words, let us assume that in the scenarios characterized by an increase of TVs (and consequently by a decrease of WAs), a 30\% of the agents, that we call the white zone, use only the information circulating in the neighborhood as their main source of information.

As one can see from the experimental settings listed in details in Table \ref{tab:expIII}, the population is composed by 100 agents (NA) endowed with the ability to process the information through a variable level of confidence (TOL). A subset of agents (WAs) can access a different source of information with values different from the ones reported on by the media (TVs).
This battery of experiments aims at understanding the effect of the two sources of information also on agents that are never directly reached by them. 
In particular the agents lying in the white zone will receive the information delivered either by experts and media but only through their neighbors. Such a process of information transmission propagates information through all the interaction topology: a message delivered by an agent will get far away on the network. 

Each scenario has been simulated in ten runs.
\begin{table}[h!]
\centering
\caption{Experiments Settings }
\begin{tabular}{|c|c|c|l|} \hline
NA&MB&WAs&T\\ \hline
100 & 0& 70&0.2,0.5 and 0.8\\ \hline
100 & 10& 60&0.2,0.5 and 0.8\\ \hline
100 & 20& 50&0.2,0.5 and 0.8\\ \hline
100 & 30& 40&0.2,0.5 and 0.8\\ \hline
100 & 40& 30&0.2,0.5 and 0.8\\ \hline
100 & 50& 20&0.2,0.5 and 0.8\\ \hline
100 & 60& 10&0.2,0.5 and 0.8\\ \hline
100 & 70& 0&0.2,0.5 and 0.8\\ \hline
\end{tabular}
\label{tab:expIII}
\end{table}

\subsubsection{Emerging Results}
\begin{figure}[h!]
 \centering
 \subfigure[Tolerance: 0.2]
   {\includegraphics[width=50mm]{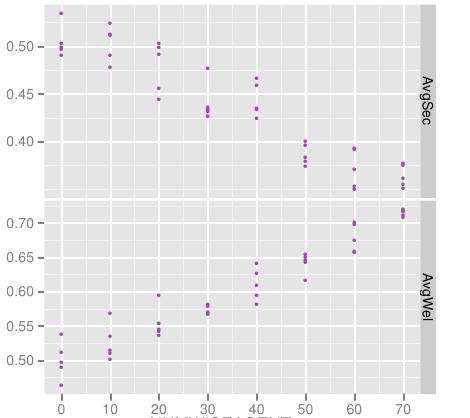} \label{fig:IIIWISE02}} 
 \hspace{1mm}
 \subfigure[Tolerance: 0.5]
  {\includegraphics[width=50mm]{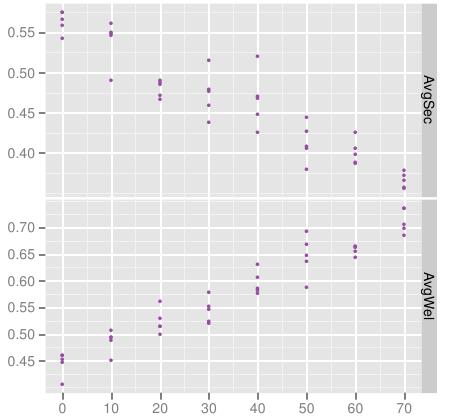} \label{fig:IIIWISE05}} 
\hspace{1mm}
\subfigure[Tolerance: 0.8]
  {\includegraphics[width=50mm]{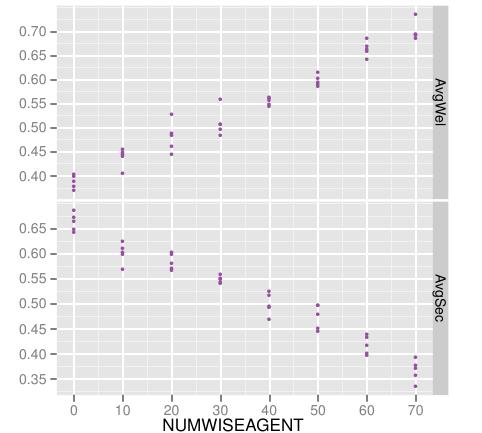} \label{fig:IIIWISE08}} 
 \caption{Opinion Trends with the 30\% of agents being not reached by media and experts}
\label{fig:expIIIP}
\end{figure}

The results emerging from this set of experiments are shown in \textbf{Figure \ref{fig:expIIIP}}, where the aggregate values of agents' opinions with respect to welfare and security are reported for the various different scenarios, each one denoted by an increasing number of WAs and a decreasing number of TVs.
The figures reported on the different scenarios stand for different values of tolerance (0.2, 0.5, 0.8). Remember that tolerance is the threshold representing the limit within which an information can be taken into account by the agents or not. Hence the higher the tolerance  the more likely the agents will be to trust informers.
As derived from our hypotheses, the white zone matters. 
Looking at the initial portion of each box in \textbf{Figure \ref{fig:expIIIP}}, in which all agents are TVs (and WAs are absents), the average values for security and welfare are never the same as the messages spread by the media. 

There is a clear evidence of the capacity of peer-to-peer communication to inhibit the effect played by the central media. Furthermore, this evidence is amplified by the white zone (i.e. the 30\% of agents not directly reached by the media nor by the experts). 
Looking at the pictures shown in \textbf{Figure \ref{fig:expIIP}}, the distance between the initial values of opinions and the media's message is smaller than the same distance in the second scenario. Looking at the pictures reproduced in \textbf{Figure \ref{fig:expIIP}}, the distance between the initial values of opinions and the media's message is smaller than the same distance in the second scenario.
Even when TVs are predominant, the central media do not lead to opinions converging on the values they transmit.

\section{Conclusions}
In this paper, the dynamics of two relatively independent opinions in a simulated network is observed. Plunged into the network, agents characterized as more or less likely to exchange opinions with neighbors are also exposed to information broadcasted by central media. 

A first series of experiments shows that when central media spread false news, P2P communication can reduce the effect of informational cheating until the broadcasting message has reached around half the population, but it tends to lose this compensating effect for increasing values of agents' exposure to informational cheating. Even a small number of experts can dramatically re-orient agents' opinions. This effect is less flashy when agents are more likely to accept others' opinions, what should not come as a surprise: the less confident agents are, the more they tend to oscillate among different opinions. Instead, the more confident they are, the lesser they are likely to converge on either the experts' or anyone else opinions. However, with a mild level of confidence the expert source is more efficacious in contrasting the impact of informational cheating.

A second series of experiments shows the effect of peer-to-peer communication. When a certain percentage of the population is not directly reached neither by the media nor by the experts, the agents' opinions do not totally converge on the messages spread by the media, not even for the highest number of agents exposed to the media and the lowest number of agents' accessing the experts. 

The results obtained so far provide some tentative answers to our initial questions. What is the impact of P2P communication on information quality, when agents are exposed to central media information flooding? It depends on how pervasive the broadcasting is: peer-to-peer communication can contrast the impact of the complementary system until when no more than 60\% of the population is reached by the broadcasted messages. How much experts must be accessed through the P2P system for contrasting informational cheating? It depends on agents' confidence in their own opinions: when this is too high or too low, a larger number of experts (30\% or more) is needed to contrast informational cheating. When confidence is neither too high nor too low, even a small percentage (around 20\% ) is enough to obtain the same results.

Is peer-to-peer communication able to inhibit the information flooding exercised by the central media, inhibiting possible information cheating and containing the corruption of information? Our simulation provides a preliminary positive answer to this question. A white zone, in which there are no televiewers nor wise agents and in which agents can access information only through their neighbors, prevents opinions from converging on the values transmitted through the central media. Peer-to-peer communication matters in reducing aggregation of opinions.

As several recent research works have outlined out the importance of the dynamic aspects of social interactions \cite{Shao09,Lorenz2008,brunetti2010,brunetti2011a,AQ2010a,AQ2010b,CFQS2010a,ACFQS2010a}, in future studies we are interested to characterise the opinions' evolution and how their behavior is affected by the dynamic nature of contacts.
In addition, we aim at implementing a more plausible model of opinions, taking into account other dimensions beside confidence, in particular the perceived correspondence between own and others' opinions and how these dimensions interact in the dynamics of beliefs.

\section{Acknowledgements}
This work was supported by the European Community under the FP6
programme (eRep project CIT5-028575). A particular thanks to Ilvo Diamanti, Federica
Mattei, Federico Cecconi, Geronimo Stilton and the Hypnotoad.
In addition we are endebted to the Italian anomaly and the Italian media for inspirations and insights.

\bibliographystyle{plain}
\bibliography{biblio}

\end{document}